\documentclass[lettersize,journal]{IEEEtran}
\usepackage{amsmath,amsfonts}
\usepackage{algorithmic}
\usepackage{array}
\usepackage[caption=false,font=normalsize,labelfont=sf,textfont=sf]{subfig}
\usepackage{textcomp}
\usepackage{stfloats}
\usepackage{url}
\usepackage{verbatim}
\usepackage{graphicx}
\def\BibTeX{{\rm B\kern-.05em{\sc i\kern-.025em b}\kern-.08em
    T\kern-.1667em\lower.7ex\hbox{E}\kern-.125emX}}
\usepackage{balance}

\usepackage{multirow}
\usepackage{graphicx}
\usepackage{booktabs} 
\usepackage{xcolor}

\begin{document}

\title{Comparative Evaluation of Static Embedding Models for
HTTP Request Anomaly Detection}


\author{\IEEEauthorblockN{Amanda Riverol} \\
\IEEEauthorblockA{Tilsor SA, ariverol@tilsor.com.uy} \\
\and
\IEEEauthorblockN{Gustavo~Betarte, Rodrigo Martínez} \\
\IEEEauthorblockA{Facultad de Ingeniería, Universidad de la República,{gustun,rodmart}@fing.edu.uy} \\
\and
\IEEEauthorblockN{Álvaro Pardo}\\
\IEEEauthorblockA{Departamento de Informática e Inteligencia Artificial,  Universidad Católica del Uruguay,apardo@ucu.edu.uy}
}

\maketitle

\begin{abstract}
Web applications are increasingly targeted by cyberattacks that exploit HTTP requests to evade security mechanisms. Traditional web application firewalls (WAFs) rely on rule-based approaches that often exhibit high false positive rates and limited adaptability. Recent studies have explored machine learning techniques and word embedding models to improve anomaly detection in HTTP traffic. This paper presents a benchmark for static embedding models—specifically Word2Vec, FastText, and Doc2Vec—within a unified, single-class classification framework. We propose HEDA (HTTP Embedding-Based Detection Architecture), a modular detection pipeline that combines static embedding representations with single-class anomaly detection models to detect anomalies at the request level. The approach operates in an unsupervised environment, where both the embedding models and detectors are trained exclusively on benign HTTP traffic. The proposed methodology is evaluated on three datasets with heterogeneous characteristics, including both synthetic and real traffic. The experimental results show that the choice of embedding representation significantly affects detection performance, and that FastText-based embeds produce the most consistent results across all datasets, achieving high detection rates while keeping false positive rates under control.

\end{abstract}


\section{Introduction}
\IEEEPARstart{T}{he}  number of web applications exposed to the Internet continues to grow steadily. These applications typically operate in a client-server architecture in which the server manages data and processes requests, while clients interact with the system via web browsers or mobile applications. From a computer security perspective, this expansion directly increases the attack surface, defined as the set of points that may be exploited by potential threats. At the same time, the volume, diversity, and complexity of HyperText Transfer Protocol (HTTP) requests have grown significantly, introducing new challenges for anomaly detection, as malicious activity must be identified within a large and heterogeneous set of legitimate traffic.

In addition, modern cyberattacks are no longer limited to static patterns or previously known vulnerabilities. Instead, adaptive attacks and variants of existing techniques have emerged, enabling adversaries to evade traditional detection mechanisms, particularly those based on fixed rules. This evolution highlights the need for more intelligent and flexible detection approaches that can adapt to evolving attack behaviors and the threat landscape.

The OWASP Top Ten project identifies the most critical security risks affecting web applications and demonstrates that these vulnerabilities remain among the primary attack vectors in practice \cite{owasp}. In response, many organizations deploy security control mechanisms such as web application firewalls (WAF). The most widely used WAF relies on the OWASP Core Rule Set (CRS) as its set of fixed rules, which focuses on identifying known attack patterns to protect web applications \cite{ghanbari2015comparative}.

Although WAFs have proven effective at mitigating a wide range of attacks, their performance heavily depends on the expertise of the security specialists who configure, tune, and maintain detection rules. Inefficient rule management can lead to elevated false positive rates (FPR), negatively affecting legitimate traffic and compromising the normal operation of protected applications.

Recent studies \cite{betarte2018web, rmartinez-ladc-2018, montes2021web} indicate that machine learning–based approaches can enhance attack detection while reducing the dependence on extensive manual rule configuration, offering a promising alternative to traditional security mechanisms. Within this context, language modeling has gained attention as a viable approach for the automatic detection of attacks in web applications, enabling the analysis of large volumes of HTTP requests and their classification based on behavioral patterns.

However, conventional text representation techniques such as one-hot encoding, Bag-of-Words (BoW), and Term Frequency-Inverse Document Frequency (TF-IDF) have limitations. These methods depend on an optimal selection of observable features and fail to capture the semantic relationships and contextual information present in HTTP requests, thereby limiting their effectiveness in detecting more sophisticated and adaptive attacks \cite{betarte2018web, rmartinez-ladc-2018}.

Advanced techniques, such as word embeddings, address these limitations by providing richer semantic representations of textual data, thereby facilitating the identification of anomalous behaviors that may evade rule-based detection systems. Nevertheless, contextualized embedding models often entail high computational costs and increased processing times. In scenarios requiring near-real-time detection, these constraints can hinder practical deployment \cite{montes2021web}.

To address these challenges, this work investigates the use of static embedding techniques, including Word2Vec, Doc2Vec, and FastText, as computationally efficient alternatives to enhance the performance of unsupervised intrusion detection models. A structured comparative evaluation of these embedding methods is performed in the context of HTTP anomaly detection to inform the design of an optimized training and inference pipeline for WAFs operating in real-world environments.

\section{Related Work}
In recent years, the effectiveness of word embedding representations has been investigated across a wide range of natural language processing tasks. Some of this research evaluates different embedding models across languages, domains, and experimental settings, providing insights into their semantic and syntactic capabilities for specific tasks. The studies summarized in Table \ref{tab:related-embedding-studies}, along with others analyzed below, illustrate how static embeddings have been evaluated and applied in contexts such as linguistic evaluation, sentiment analysis, and text classification, offering a methodological basis for exploring their use in more specialized domains, such as the one addressed in this article: anomaly detection in web applications.

\subsection{Static Embedding Models in Natural Language Processing}

Bulbul et al. \cite{bulbul2024comprehensive} present a comprehensive evaluation of static word embedding models, including Word2Vec, FastText, and GloVe, trained in a large Bengali newspaper corpus. By analyzing both the Continuous Bag-of-Words (CBOW) and Skip-Gram architectures and comparing models trained from scratch with those generated using Gensim, the study shows that Skip-Gram generally outperforms CBOW. FastText achieves superior results on syntactic tasks, while Word2Vec excels in semantic analogies. The authors also demonstrate that embedding quality is strongly influenced by training methodology and hyperparameter selection. Although the study focuses on the Bengali language, its evaluation framework is transferable to other languages and domains.

Pak et al. \cite{pak2024survey} provide an extensive survey of word embedding techniques, ranging from classical methods such as LSA (Latent Semantic Analysis) and LDA (Latent Dirichlet Allocation) to static neural embeddings (Word2Vec, GloVe, FastText) and modern contextual models (BERT, GPT). Through intrinsic and extrinsic evaluations across multiple NLP tasks, the authors highlight the limitations of static embeddings in handling context and polysemy, while emphasizing the superior semantic capabilities of contextual models. Although centered on general NLP tasks, the taxonomy and methodological insights are relevant to domain-specific applications such as anomaly detection.

Vianna et al. \cite{vianna2024sentiment} evaluate sentiment analysis models on Portuguese Twitter data, comparing traditional machine learning approaches based on Bag-of-Words and TF-IDF with deep learning models based on static embeddings. Their results indicate that FastText performs best on short, noisy text, underscoring the importance of language-specific resources and preprocessing for embedding-based models.

Siagh et al. \cite{siagh2023embedding} analyze the impact of embedding representations and data quality on sentiment classification using educational social media datasets. By comparing contextual embeddings (BERT) with static embeddings under varying preprocessing conditions, they show that static embeddings perform better on cleaned text, whereas BERT excels on raw input. These findings emphasize that embedding effectiveness depends strongly on input structure and preprocessing, a consideration aligned with our comparative analysis.

Dharma et al. \cite{dharma2022cnn} conduct a comparative study of Word2Vec and FastText embeddings applied to news classification using a CNN architecture. FastText achieves the highest accuracy, attributed to its subword modeling capabilities, which improve handling of out-of-vocabulary (OOV)  terms. While the study confirms the effectiveness of static embeddings in text classification, it does not explore how input structure or application context, such as anomaly detection, influences performance.

\begin{table*}[!t]
    \caption{Related Work: Comparing Static Word Embeddings}
    \label{tab:related-embedding-studies}
    \centering
    \footnotesize
    \begin{tabular}{| p{2.5cm} | p{3.2cm}| p{3.2cm}| p{7cm}|}
        \hline
        \textbf{Author(s)} & \textbf{Embeddings Compared} & \textbf{Task / Domain} & \textbf{Main Findings} \\
        \hline
        Bulbul et al.~\cite{bulbul2024comprehensive} &
        Word2Vec, FastText, GloVe &
        Bengali Language, Analogy Tasks &
        Skip-gram outperforms CBOW; FastText is best for syntax; training methodology and hyperparameters significantly impact embedding quality. \\
        \hline
        
        Pak et al.~\cite{pak2024survey} &
        Word2Vec, FastText, GloVe, BERT, GPT &
        General NLP &
        Survey and taxonomy; contextual embeddings outperform static ones in polysemy and semantics; highlights evaluation practices. \\
        \hline
        
        Vianna et al.~\cite{vianna2024sentiment} &
        Word2Vec, FastText, GloVe &
        Portuguese Tweets, Sentiment Analysis &
        FastText yields best performance on short, noisy text; language-specific resources and preprocessing are crucial. \\
        \hline
        
        Siagh et al.~\cite{siagh2023embedding} &
        Word2Vec, FastText, GloVe, BERT &
        Social Media, Education (Sentiment Classification) &
        Static embeddings perform better on preprocessed data; BERT excels on raw input; embedding effectiveness depends on input quality. \\
        \hline
        
        Dharma et al.~\cite{dharma2022cnn} &
        Word2Vec, FastText, GloVe &
        News Classification (CNN) &
        FastText achieves the highest accuracy (97.2\%); OOV handling via subwords is advantageous. \\
        \hline
        
        Kumar et al.~\cite{kumar2025analysis} &
        Word2Vec, FastText, GloVe (Ensemble) &
        Product Reviews, Sentiment Analysis &
        Vector ensemble enhances semantic robustness; applied to Amazon reviews with SVM classifier. \\
        \hline
        
        Habbat \& Nouri~\cite{nassera2025investigation} &
        Word2Vec, FastText, GloVe, Doc2Vec, BERT, GPT &
        Topic Modeling &
        Contextual embeddings improve topic coherence; static embeddings remain useful when adapted to task/domain. \\
        \hline
    \end{tabular}
\end{table*}

Kumar et al.\cite{kumar2025analysis} proposed a sentiment analysis that uses static word embeddings Word2Vec and FastText for keyword-based filtering and classification of product reviews. Their approach employs an ensemble model that averages vectors from all three embedding types to enhance semantic coverage and robustness. When applied to the Amazon Fine Food Reviews dataset, the system enables fine-grained sentiment analysis by focusing on specific product aspects (e.g., price, quality) and using an SVM classifier. 

Habbat and Nouri \cite{nassera2025investigation} provide an in-depth review of word embedding techniques used in topic modeling tasks. The study distinguishes between static word embeddings (e.g., Word2Vec, FastText, and Doc2Vec) and contextualized embeddings (e.g., ELMo, BERT, GPT, and XLNet). While static embeddings generate fixed word representations that ignore context, contextual embeddings adapt to the surrounding text, capturing richer semantic meaning. The authors emphasize that contextualized embeddings significantly enhance topic coherence, especially when combined with neural topic models based on variational autoencoders. The review also highlights the importance of selecting and fine-tuning embeddings based on the task and domain to improve topic modeling performance.

Building on the use of static embedding models in general NLP tasks, recent studies have begun to explore their application in more specialized domains. In particular, static embedding representations have been adapted to model HTTP requests as text for anomaly detection in web applications.

\subsection{Static Embedding Models for HTTP Anomaly Detection}
Although no prior comparative study has evaluated how static word embeddings affect anomaly detection effectiveness, several works have explored their application in this area, as discussed below.

Previous work \cite{rmartinez-ladc-2018,betarte2018web} has explored the application of machine learning techniques to enhance WAFs for attack detection. The problem is typically formulated under different learning scenarios depending on the availability of benign and attack data, including multi-class settings when both types of data are present and one-class approaches when only benign samples are available. In these studies, classical text representations based on TF-IDF with expert-selected tokens are commonly employed and have shown competitive performance, even in unsupervised or semi-supervised scenarios, outperforming rule-based WAF configurations such as ModSecurity with the OWASP Core Rule Set. However, these approaches are highly dependent on expert knowledge for feature selection, underscoring the need to explore more expressive and automated representation-learning techniques to reduce expert bias and improve generalization.

Li et al. \cite{li2020weightedword2vec} propose an efficient anomaly detection model for HTTP traffic based on a combination of Word2Vec and TF-IDF. Their method constructs paragraph vectors by weighting word embeddings with TF-IDF scores, thereby capturing semantic and contextual relevance. The approach is evaluated using three public datasets (HTTP DATASET CSIC 2010, UNSW-NB15, and Malicious-URLs) and demonstrates excellent performance when combined with boosting-based classifiers such as LightGBM and CatBoost. Although the study confirms the effectiveness of TF-IDF-weighted Word2Vec representations for HTTP traffic classification, it does not investigate the influence of other alternative embedding models.

Montes et al. \cite{montes2021web} apply deep learning techniques to improve WAF performance by modeling HTTP requests as text. A pre-trained, transformer-based language model (RoBERTA) is used as a feature extractor. RoBERTA transforms the requests into feature vectors, which are then used to classify them using a one-class support vector machine (OCSVM). These approaches eliminate the need for expert-defined features but are often computationally expensive.

\begin{table*}[!t]
    \caption{Related Work: Static Word Embeddings for Anomaly Detection in Web Requests}
    \label{tab:embedding_studies}
    \centering
    \footnotesize
    \begin{tabular}{| p{1.9cm}| p{2cm}| p{2cm}| p{2cm}| p{2.0cm}| p{5.2cm}|}
        \hline
        \textbf{Author(s)} &
        \textbf{Embedding} &
        \textbf{Feature Representation} &
        \textbf{Classifier / Model} &
        \textbf{Dataset(s)} &
        \textbf{Main Contribution} \\
        \hline
        
        Aharon et al.~\cite{aharon2025classification} &
        FastText (custom) &
        API token embeddings &
        ANN retrieval (few-shot) &
        CSIC 2010, ATRDF 2023 &
        Unsupervised anomaly detection using custom FastText and approximate nearest neighbors. \\
        \hline
        
        Yamada \& Kawahara~\cite{yamada2024evaluation} &
        FastText &
        Token embeddings from HTTP &
        CNN Autoencoder &
        CSIC 2010 &
        Demonstrated that FastText outperforms ASCII input in the WAF context. \\
        \hline
        
        S. Amutha et al.~\cite{amutha2024employing} &
        Doc2Vec &
        Document-level vectors &
        RF, KNN &
        Synthetic XSS Dataset &
        Effective semantic feature extraction for XSS with explainability. \\
        \hline
        
        Wang et al.~\cite{wang2023semantic} &
        Word2Vec &
        Averaged sentence embeddings &
        Logistic Regression, XGBoost &
        Kaggle SQLi dataset &
        Semantic-aware detection of SQLi using classical ML. \\
        \hline
        
        Montes et al.~\cite{montes2021web} &
        RoBERTa &
        Contextual embeddings &
        SVM &
        CSIC 2010, DRUPAL, PKDD &
        Combines semantic embeddings with term-weighting. \\
        \hline
        
        Khalil et al.~\cite{gniewkowskihttp2vec} &
        Doc2Vec, RoBERTa &
        Request-level embeddings &
        SVM &
        CSIC 2010 &
        Proposed HTTP2Vec for embedding complete HTTP requests. \\
        \hline
        
        Li et al.~\cite{li2020weightedword2vec} &
        Word2Vec + TF-IDF &
        Method, host, path, parameters &
        CatBoost &
        CSIC 2010 &
        Employs contextualized semantic embeddings to enhance feature representation\\
        \hline
        
        Betarte et al.~\cite{betarte2018web} &
        TF-IDF &
        Selected tokens &
        SVM &
        CSIC 2010, DRUPAL, PKDD &
        Traditional baseline using lexical features for anomaly detection. \\
    \hline
    
    \end{tabular}
\end{table*}

Yamada and Kawahara \cite{yamada2024evaluation} proposed an unsupervised anomaly detection for web application firewalls based on two autoencoder models. The first model encodes HTTP requests as ASCII values, while the second uses FastText to generate word embeddings, which are then processed by a convolutional autoencoder. Experiments on the CSIC2010 dataset showed that a FastText-based model significantly outperformed an ASCII-based model. The results demonstrate that incorporating semantic-aware static embeddings, such as FastText, improves detection accuracy without requiring labeled attack data. 

In \cite{amutha2024employing}, they employ Doc2Vec embeddings to extract semantic features from web request data to detect Cross-Site Scripting (XSS) attacks. The authors preprocess the input by tokenizing, normalizing, and decoding it and then use Doc2Vec to generate fixed-length vector representations of the requests. These embeddings are used to train K-Nearest Neighbors (KNN) and Random Forest (RF) classifiers. The experimental results show that the Doc2Vec-based feature representation allows the Random Forest model to achieve a detection accuracy of 97.4\%. The paper demonstrates that static document-level embeddings, such as Doc2Vec, can be effectively leveraged to detect XSS attacks and that they also enable model explainability through feature-importance analysis and tree visualization.

Aharon et al.\cite{aharon2025classification} proposed an unsupervised few-shot anomaly detection to identify API injection attacks. Their approach combines a custom FastText embedding model trained on API traffic with an Approximate Nearest Neighbor (ANN) search in a classification-by-retrieval strategy. The system was evaluated on the CSIC 2010 and ATRDF 2023 datasets, achieving high accuracy while remaining efficient and scalable. They introduced a dedicated tokenizer tailored for API request structure and emphasized the use of domain-agnostic models.

These studies, along with other related research, are summarized in Table \ref{tab:embedding_studies} and show that, in recent years, the literature has tended to adopt static embedding representations for HTTP request vectorization in the context of anomaly detection in web applications. When combined with various classification algorithms, these representations have demonstrated robust performance and high accuracy across diverse datasets. However, the analysis also reveals a lack of consistency in the evaluation process, as different embedding methods and learning algorithms are used across heterogeneous experimental conditions. Consequently, the literature lacks a systematic comparison of these representations across equivalent HTTP request environments, as is the case in other widely studied contexts, such as sentiment analysis or general natural language processing, as illustrated in Table \ref{tab:related-embedding-studies}. This deficiency makes it difficult to determine which embedding approach best represents the normal behavior of web applications and how this choice impacts the performance of anomaly detection models.

\section{Static Embedding Models}
Machine learning based on natural language processing (NLP) has advanced significantly in recent years. A fundamental component of this process is feature extraction, which involves converting textual data into numerical representations, commonly known as word embeddings or vector representations.

Early embedding models, such as Word2Vec, generate static representations, meaning that each word is assigned a single vector regardless of its context. Word2Vec is based on a neural network that implements two architectures: CBOW and Skip-Gram, to learn relationships between words from local context windows \cite{mikolov2013efficient}.

Doc2Vec \cite{le2014distributed} is an extension of Word2Vec that learns global representations of text, such as sentences or documents, capturing broader semantic information. However, it still produces static vectors and often offers only modest improvements over simple word embedding.

FastText\cite{bojanowski2017enriching}, another extension of Word2Vec, incorporates subword information using character-level n-grams, allowing for better handling of uncommon, morphologically complex, or unseen words in the training set. Despite this improvement, FastText also produces static embeds and cannot fully capture contextual meaning.

Recent advances in NLP have introduced contextualized embedding models that generate word representations that vary with context. Unlike static embeddings such as Word2Vec, Doc2Vec, or FastText, which associate each word with a single fixed vector, contextual models produce dynamic representations that capture semantic differences between sentences. Prominent examples include transformer-based architectures, such as BERT \cite{seyyar2022attack} and RoBERTa, which leverage deep neural networks to model bidirectional contextual dependencies in text. This capability allows contextual embeddings to better represent polysemy and subtle semantic relationships that static embeddings cannot capture. However, these models typically require more computational and memory resources and introduce higher inference latency ~\cite{aharon2025classification}. As a result, despite their greater representational capacity, contextualized embeddings can have limitations in real-time or resource-constrained applications, such as the one presented in this article, where faster and lighter static embedding methods may still be preferable.

\subsection{Selected Static Embedding Models}
In word embedding approaches, each word is represented as a dense vector in a continuous vector space whose dimensionality is significantly smaller than the vocabulary size. These representations capture semantic and syntactic relationships between words, allowing machine learning models to process textual data more effectively. Different embedding models generate these vector representations using distinct training strategies and objectives, resulting in variations in the types of linguistic information they capture.

In this work, several static word embedding models are considered in order to compare their effectiveness for representing textual information in the context of web application attack detection. The selected models are described in the following subsections.

\subsubsection{\textbf{Word2Vec}} 

Word embedding model introduced by Mikolov et al. \cite{mikolov2013efficient} maps each word in a text into a continuous vector representation using an unsupervised neural network. The model consists of a projection (hidden) layer and is trained via stochastic gradient descent and backpropagation. Words that frequently appear within similar context windows tend to share similar vector representations due to shared weights.

\begin{figure}[!t]
    \centering
    \includegraphics[width=1.8in]{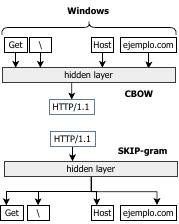}
    \caption{Word2Vec architectures: CBOW and Skip-Gram.}
    \label{fig:embedding-word2Vec}
\end{figure} 

Word2Vec includes two main architectures: CBOW and Skip-Gram. CBOW predicts a target word based on its surrounding context, while Skip-Gram predicts surrounding words from a target word (Figure~\ref{fig:embedding-word2Vec}). Both models rely on a sliding window to define the local context and are capable of capturing syntactic and semantic relationships between tokens

The performance of Word2Vec is highly dependent on the hyperparameter selection, including:

\begin{itemize}
    \item \textbf{Architecture:} Defines the training algorithm. \texttt{sg=0} corresponds to CBOW (predicts target word from context), while \texttt{sg=1} selects Skip-Gram (predicts context words from a target word).
    \item \textbf{Vector Size:} Dimensionality of the embedding space. 
    \item \textbf{Context Window:} Size of the context window. A window of 2 considers two words before and after the target word.
\end{itemize}

In the context of HTTP anomaly detection, Word2Vec is typically applied to tokens extracted from HTTP requests, including method names, URL paths, query parameters, and header values. By learning co-occurrence patterns from benign traffic, the model can encode typical request structures. However, because Word2Vec assigns a single vector per token and ignores morphological structure, it is sensitive to vocabulary size and is disadvantaged by rare, malformed, or obfuscated tokens, which are common in malicious requests.

\subsubsection{\textbf{Doc2Vec:}} Known as Paragraph Vector, is an extension of Word2Vec proposed by Le and Mikolov \cite{le2014distributed}. It learns fixed-length vector representations for larger text units such as sentences, paragraphs, or entire documents. Each document is associated with a unique vector trained jointly with the words it contains, allowing the model to capture broader semantic information beyond individual tokens.

\begin{figure}[!t]
    \centering
    \includegraphics[width=2.5in]{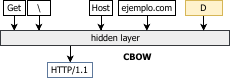}
    \caption{Doc2Vec architecture.}
    \label{fig:embedding-doc2Vec}
\end{figure}

Doc2Vec provides two main architectures: Distributed Memory (DM) (Figure \ref{fig:embedding-doc2Vec}), which incorporates context words and a document identifier to predict a target word, and Distributed Bag of Words (DBOW), which predicts words directly from the document vector.

When applied to HTTP traffic, Doc2Vec allows the representation of entire HTTP requests as single vectors, capturing global patterns. This property makes it suitable for anomaly detection approaches that operate at the request level. However, as Word2Vec, the learned representations are static and tightly tied to the distribution observed during training, which can limit their ability to generalize to previously unseen attack patterns. Furthermore, subtle or localized anomalies can be diluted within the global representation of the request, thereby reducing their detectability.

\subsubsection{\textbf{FastText:}} It's an extension of the Word2Vec model, but it represents words as sets of $n$-grams at the character level. Each $n$-gram is associated with a vector, and the word's embedding is calculated as the sum of the embeddings of the $n$-grams that constitute it. The boundary symbols \text{<} and \text{>} are added to distinguish prefixes and suffixes from internal character sequences.

For example, given the token \textbf{\textit{HTTP/1.1}} and $n = 3$, FastText generates the following $n$-grams of characters:

    \begin{center}
        \texttt{<HT, HTT, TTP, TP/, P/1, /1., 1.1,.1>}
    \end{center}
    
    along with the full token:
    
    \begin{center}
        \texttt{<HTTP/1.1>}
    \end{center}

In the context of FastText, the model is particularly effective at detecting HTTP anomalies, thanks to its ability to model subword patterns presented in the request. By leveraging character-level information, the model can generate meaningful representations for OOV and obfuscated tokens, which are common in malicious traffic. This capability can improve robustness in dynamic and noisy environments.

\section{Unsupervised Learning Methodology}

This comparative study focuses on the role of the representation stage in the performance of request-level anomaly detection systems. In this context, different static embedding-based representations are analyzed to identify the model that best captures the characteristics of HTTP traffic. Although embedding models have been extensively evaluated on traditional natural language processing tasks (see Table~\ref{tab:related-embedding-studies}), HTTP requests exhibit domain-specific properties that differ substantially from those of natural language, motivating a dedicated evaluation.

\begin{figure*}[!t]
\centering
  \includegraphics[width=5.5in]{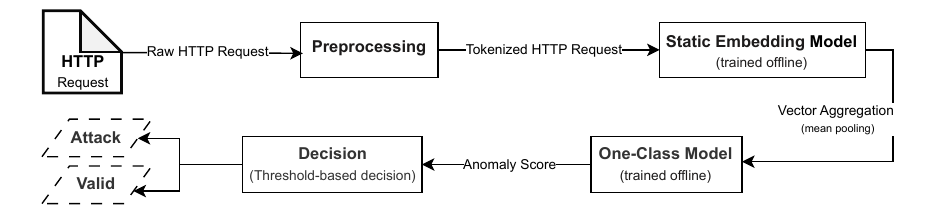}
  \caption{HEDA: HTTP Embedding-based Detection Architecture.}
  \label{fig:heda-scheme}
\end{figure*}

To support this analysis, a modular detection pipeline is proposed that enables request-level anomaly detection by processing each HTTP request independently, as shown in Fig. \ref{fig:heda-scheme}. This architecture allows for a controlled and fair comparison of both embedding representations and anomaly detection algorithms.

The proposed pipeline is conceptually divided into two main stages, with particular emphasis on the representation stage. In this stage, different static embedding models can be instantiated while keeping the remaining components fixed. This design ensures that any observed performance differences can be directly attributed to the representation model under evaluation.

The proposed pipeline assumes the availability of a trained static embedding model. Pretrained embeddings are available for natural language processing tasks, however, this work incorporates a dedicated offline training phase tailored to the HTTP domain. This design choice aims to capture meaningful relationships involving special characters and tokens, which often act as strong indicators of malicious behavior in HTTP requests but are typically ignored in conventional NLP settings.

FastText, Word2Vec, and Doc2Vec models are trained using multiple configurations 
, with embedding dimensions ranging from 50 to 200 and context window sizes of 3, 5 and 7. This setup enables a systematic comparison across both embedding models and hyperparameter choices.

Once the embedding training stage is complete, Phase 2 
is executed, in which an anomaly detection model is trained for each generated embedding model. For each execution, an anomaly detection model is trained using the representations generated by the embedding model trained in that instance.

The detection model operates under a one-class learning paradigm and employs an operational point selection procedure to determine the anomaly score threshold for inference.

At the end of this phase, multiple one-class anomaly detection models are generated from different static embedding-based representations. Each model will be evaluated to identify the one that achieves the highest recall when combined with the classification algorithm, while maintaining a low FPR.

Emphasis is placed on recall due to the security-sensitive nature of the problem, where undetected attacks can have serious consequences, but without incurring in a high number of false positives that could negatively affect the system's usability.

\section{Experimental Results}
The primary objective of this study is to analyze the impact of employing various static embedding representations on the performance of anomaly detection in HTTP requests.

In the initial phase, various static embedding representations were assessed while maintaining the anomaly detection algorithm (OCSVM) unchanged. Specifically, the objective was to determine the embedding configuration that most effectively captures the statistical structure of legitimate HTTP traffic and is most appropriate for subsequent anomaly detection.

During the second phase of experimentation, several one-class anomaly detection models were evaluated using a uniform static embedding representation. This investigation assesses the efficacy of the OCSVM model relative to other one-class anomaly detection techniques, all trained on identical static embeddings of HTTP traffic.

The experiments followed an identical preprocessing sequence and used a consistent data partitioning strategy. Both the embedding and anomaly-detection models were exclusively trained on legitimate HTTP requests. A validation subset was used for operating-point calibration, and the final performance was assessed on a test set comprising both legitimate and malicious traffic.

\subsection{HTTP Request Preprocessing}
\label{sec:http-preprocessing}

Each request is converted to a text string and decoded twice using URL decoding to recover characters that may have been encoded multiple times. This step helps reveal patterns that might be hidden in the request's encoding. After decoding, the request is converted to lowercase to ensure case-insensitive processing and reduce variability in the text representation (proposed in~\cite{rmartinez:thesis} and subsequently employed in~\cite{rmartinez-ladc-2018,montes2021web,riverol2024capturing}). 

The request is then tokenized using the NLTK tokenizer. This tokenizer splits the request into a sequence of tokens based on punctuation and whitespace. As a result, each valid HTTP request in the dataset is represented as an ordered sequence of tokens, which serves as input for training the embedding models used in the proposed framework.

\subsection{Selection of Embedding Models and Hyperparameter Configurations}
The experiments in this section use the Drupal dataset, which was introduced in~\cite{rmartinez-ladc-2018}. This dataset consists primarily of benign HTTP requests and reflects real-world deployment conditions. Training is performed exclusively using valid traffic, while a set of attack requests is reserved for testing and evaluation. Table~\ref{tab:dataset} summarizes the main characteristics of the dataset.

\begin{table*}[!t]
\caption{Statistics of Dataset}
\label{tab:dataset}
\centering
\footnotesize
    \begin{tabular}{| l | l | c | c | c | c | c | c | c | c | c |}
    \hline
        \multirow{2}{*}{\textbf{Split}} &
        \multirow{2}{*}{\textbf{Class}} &
        \multicolumn{3}{c|}{\textbf{DRUPAL}} &
        \multicolumn{3}{c|}{\textbf{CSIC 2010}} &
        \multicolumn{3}{c|}{\textbf{SR-BH 2020}} \\
         \cline{3-11}
             & &
            \textbf{Req} & \textbf{Words} &\textbf{Uniq} 
            & \textbf{Req} & \textbf{Words} &\textbf{Uniq}
             &\textbf{Req} & \textbf{Words} &\textbf{Uniq} 
            \\
         \hline
         Train & Valid   & 45\,609 & 6\,425\,269 & 13\,972  
                         & 36\,000 &7\,117\,978 &58\,055  
                         &367\,547&43\,702\,525&12\,560 \\ 
        \hline
         \multirow{2}{*}{Test} 
         & Valid   & 19\,533 & 2\,745\,495 & 8\,777  
                    &36\,000 &7\,118\,738 & 58\,744 
                    &157\,648&18\,752\,952&7\,146 \\ 
         & Attacks & 325     & 44\,884     & 573     
                   & 25\,065&5\,498\,727 &40\,821 
                    &114\,697&19\;045\,348&35\,879 \\
         \hline
    \end{tabular}
\end{table*}

This study evaluates three widely adopted static embedding methodologies for textual representation: Word2Vec, Doc2Vec, and FastText. Each model undergoes offline training exclusively utilizing valid HTTP request strings from the dataset. Those are represented as sequences of tokens, as delineated in Section~\ref{sec:http-preprocessing}. The resulting embeddings are subsequently employed as inputs for the anomaly detection models. The models' performance is compared using the following metrics: FPR, True Positive Rate (TPR), and Area Under the Curve (AUC).

To analyze the effect of representational capacity and contextual scope, multiple embedding configurations were explored by varying the embedding dimension $d \in \{50, 100, 150, 200\}$, the context window size $w \in \{3, 5, 7\}$, and the embedding model (Word2Vec, FastText, and Doc2Vec). This results in 36 configurations, of which 12 are evaluated for each embedding model. Preliminary experiments with larger embedding dimensions did not yield significant performance improvement.

According to the statistics of the evaluated datasets, the number of tokens per HTTP request ranges from a few tens to several hundred tokens. For instance, in the Drupal dataset, requests range from 8 to 573 tokens, with an average of approximately 141 tokens, whereas the CSIC and SR-BH datasets have average lengths of 119-198 tokens. Most requests fall within the 100-200 token range, indicating that the selected embedding configurations are sufficient to capture relevant contextual information within typical request lengths.

\begin{table}[!t]
\caption{Performance of the One-Class SVM Using Different Static Embedding Configurations With Optimized Decision Thresholds}
\label{tab:raw_optim_results}
\centering
\footnotesize
    \begin{tabular}{| l | c | c | c | c | c |}
        \hline
        \multicolumn{6}{|c|}{\textbf{DRUPAL}} \\
        \hline
        \textbf{Embedding} & \textbf{$d$} & \textbf{$w$} &
        \textbf{FPR (\%)} & \textbf{TPR (\%)} & \textbf{AUC} \\
        \hline
          \multirow{9}{*}{\textbf{Word2Vec}} &
          \multirow{3}{*}{100}&
              3 & 0.35 & 70.26 & 0.82 \\
          & & 5 & 0.40 & 71.40 & 0.85 \\
          & & 7 & 0.36 & 71.40 & 0.88 \\
         
          &\multirow{3}{*}{150}&
            3 & 0.27 & 68.37 & 0.83 \\
            &  & 5 & 0.36 & 70.45 & 0.85 \\
            &  & 7 & 0.45 & 72.53 & 0.87 \\
        
         &\multirow{3}{*}{200}&
            3 & 0.28 & 68.75 & 0.82 \\
            &  & 5 & 0.29 & 68.56 & 0.86 \\
            &     & 7 & 0.29 & 70.83 & 0.88 \\
        \hline

        \multirow{9}{*}{\textbf{FastText}} &
         \multirow{3}{*}{100}&
              3 & 0.95 & 81.11 & 0.93 \\
             &  & 5 & 0.75 & 80.11 & 0.95 \\
             &  & 7 & \textbf{0.79} & \textbf{80.11} & 0.96 \\
        
                 &  & 3 & 0.99 & 80.30 & 0.94 \\
          & 150    & 5 & 0.98 & 80.87 & 0.95 \\
                 &     & 7 & 0.75 & 80.49 & 0.96 \\
        
                 &  & 3 & 1.04 & 80.68 & 0.93 \\
         & 200 & 5 & 0.79 & 80.68 & 0.95 \\
                 &     & 7 & 0.77 & 80.87 & 0.96 \\
        \hline

        \multirow{9}{*}{\textbf{Doc2Vec}} &
        \multirow{3}{*}{100}&
           3 & 0.50 & 54.54 & 0.89 \\
           & & 5 & 0.41 & 52.65 & 0.86 \\
           & & 7 & 0.39 & 55.68 & 0.87 \\
        
        & \multirow{3}{*}{150}&
            3 & 0.52 & 54.73 & 0.90 \\
            &  & 5 & 0.35 & 54.73 & 0.89 \\
            &  & 7 & 0.23 & 53.97 & 0.87 \\
        
        & \multirow{3}{*}{200}&
            3  & 0.51 & 56.06 & 0.90 \\
            &  & 5 & 0.27 & 50.18 & 0.87 \\
            &  & 7 & 0.18 & 51.13 & 0.87 \\
        \hline
    
    \end{tabular}
\end{table}

Table~\ref{tab:embedding_coverage} compares the vocabulary coverage of the evaluated embedding models. FastText achieves complete token coverage thanks to its subword modeling mechanism, which enables it to generate representations for previously unseen tokens using character n-grams. 

\begin{figure*}[t]
    \centering
    \includegraphics[width=1\textwidth]{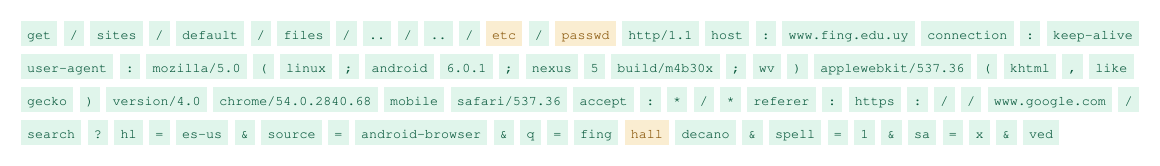}
    \label{fig:fasttext}
    \caption{FastText generates representations for OOV tokens using subword n-grams.}
    \label{fig:embedding-coverage}
\end{figure*}

Figure~\ref {fig:embedding-coverage} demonstrates an example from the dataset in which an HTTP request contains tokens that were not encountered during the training phase. In such instances, FastText is capable of generating vector representations utilising subword components (tokens highlighted in yellow), whereas Word2Vec can only depict tokens explicitly present in the training vocabulary. Consequently, as shown in Table~\ref{tab:embedding_coverage}, FastText encompasses the entire valid and attack request, whereas Word2Vec accounts for only 85.1\% of valid and 6.6\% of attack requests.

\begin{table}[t]
    \centering
    \caption{Vocabulary coverage of embedding models on the test set}
    \label{tab:embedding_coverage}
    \footnotesize
    \begin{tabular}{| l | c | c | c|}
        \hline
        \textbf{Metric (vocabulary=16,488)} & \textbf{Doc2Vec} & \textbf{FastText} & \textbf{Word2Vec} \\
        \hline
        Total token coverage (test) & 99.0\% & 100.0\% & 99.0\% \\
        Direct vocabulary coverage & 99.0\% & 99.0\% & 99.0\% \\
        Coverage via n-grams & 0.0\% & 1.0\% & 0.0\% \\
        OOV tokens without vector & 1.0\% & 0.0\% & 1.0\% \\
        \hline
        Requests with representation & 100.0\% & 100.0\% & 100.0\% \\
        Fully covered requests & 83.1\% & 100.0\% & 83.1\% \\
        Fully covered valid requests& 85.1\% & 100.0\% & 85.1\% \\
        Fully covered attack requests& 6.6\% & 100.0\% & 6.6\% \\
        \hline
    \end{tabular}
\end{table}

The performance of the OCSVM classifier depends on two hyperparameters: $\nu$ and $\gamma$. The parameter $\gamma$ defines the kernel coefficient of the radial basis function (RBF) kernel and controls the shape of the decision boundary, while $\nu$ specifies an upper bound on the fraction of training samples considered as outliers and a lower bound on the fraction of support vectors \cite{6790022}. For each embedding configuration, an OCSVM is trained using only valid requests from the training set. The OCSVM hyperparameters are selected from predefined discrete sets, with $\nu \in \{0.001, 0.05, 0.1, 0.5\}$ and $\gamma \in \{0.001, 0.1, 0.5\}$. 

After training, an operating point selection process is performed to calibrate each detector's decision threshold. The anomaly decision thresholds are swept over the range $[-25, 25]$ with a step size of $0.01$.  The optimal configuration is determined through the $\hat{F}$ measure proposed in~\cite{lee2003learning} and used in~\cite{montes2021web}, which can be estimated using normal and unlabeled examples and exhibits behavior similar to the $F_1$-score. The parameter combination that maximizes $\hat{F}$ on a validation set composed of normal and unlabeled requests is selected.

\begin{figure*}[!t]
\centering    
    \subfloat[\small ]{
    \includegraphics[width=0.30\textwidth]{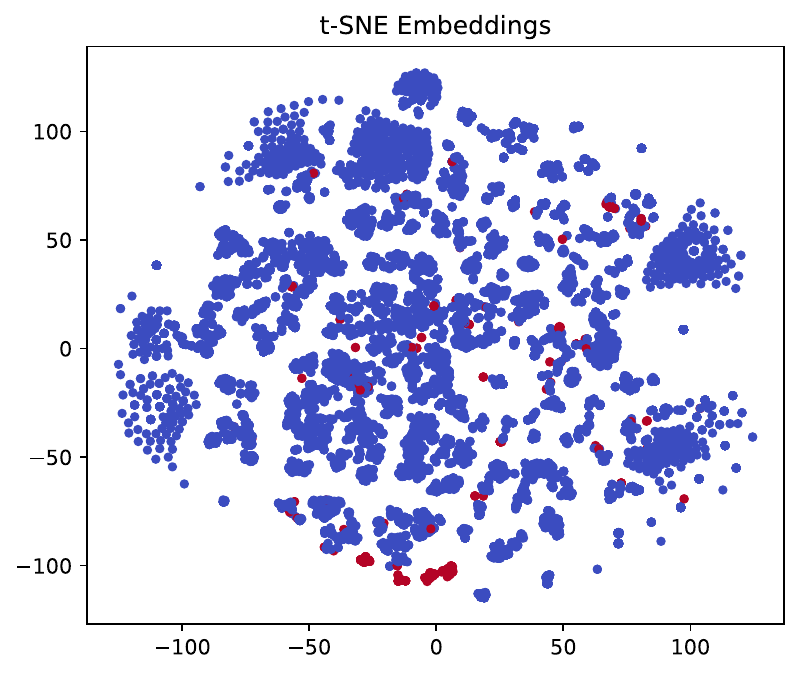}
    \label{fig:tsne_embeddings}}
    \hfill
    \subfloat[\small ]{
    \includegraphics[width=0.25\textwidth]{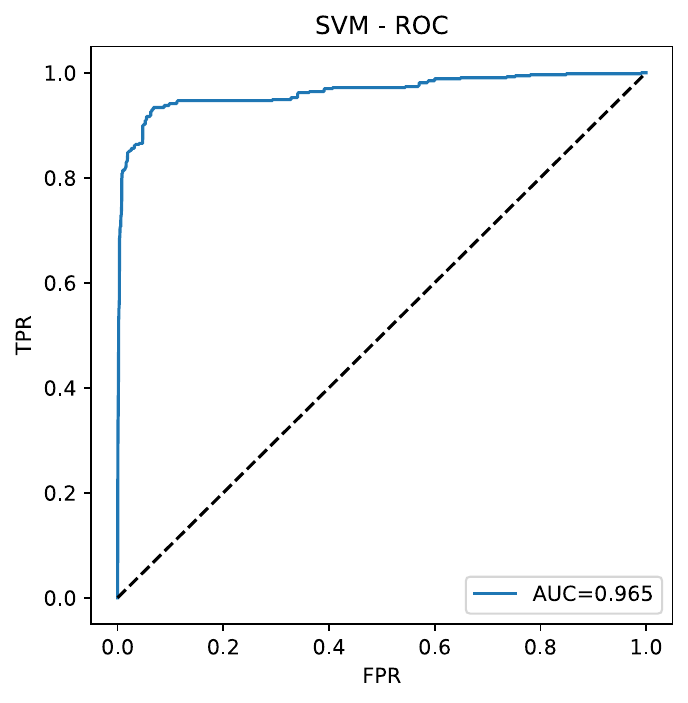}
    \label{fig:svm_roc}}
    \hfill
    \subfloat[\small ]{
    \includegraphics[width=0.35\textwidth]{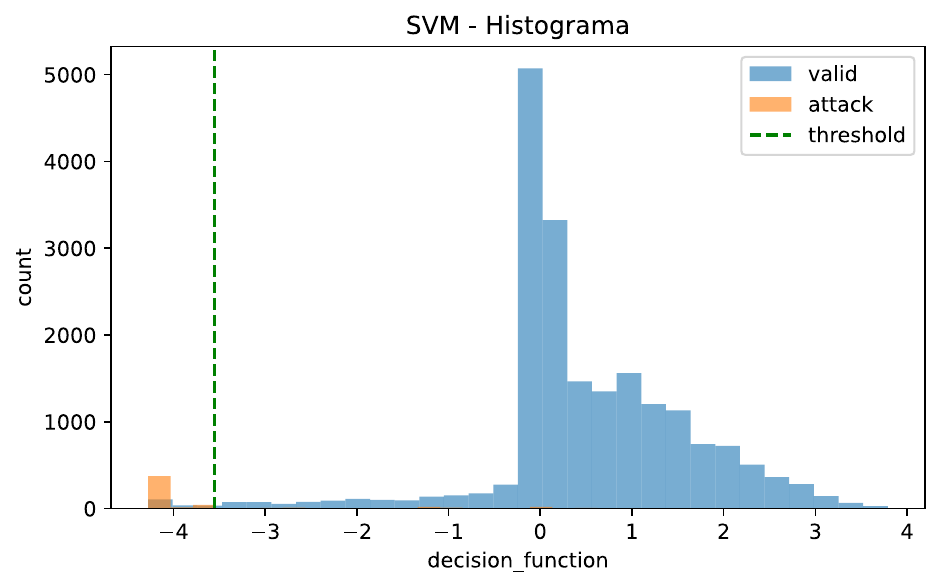}
    \label{fig:svm_hist}}
    
    \caption{Results on the Drupal dataset using FastText. (a) t-SNE projection (blue: valid, red: attack), (b) ROC curve of the OCSVM, and (c) distribution of anomaly scores.}
\label{fig:embedding_ocsvm_results}
\end{figure*}

Table~\ref{tab:raw_optim_results} reports the performance of the OCSVM across the different embedding configurations. Overall, FastText-based representations consistently outperform Word2Vec and Doc2Vec when combined with the OCSVM, achieving the highest AUC and TPR values across the evaluated configurations. The learned embedding space for FastText is illustrated in Fig.~\ref{fig:tsne_embeddings}, which provides a two-dimensional visualization of the test dataset's representations.

Among the evaluated configurations, the model with embedding dimension $d = 100$ and context window size $w = 7$ provides the best balance between detection performance and stability, achieving an AUC of 0.96, a TPR above 80\%, and a low FPR, as shown in Fig.~\ref{fig:svm_roc}. Additionally, Fig.~\ref{fig:svm_hist} shows the distribution of anomaly scores produced by the detector, highlighting a clear separation between normal and anomalous requests.

Since higher-dimensional configurations do not yield significant performance improvements, FastText with parameters ($d = 100$, $w = 7$) is selected as the final embedding configuration due to its strong detection performance and lower computational complexity. To analyze the generalization capability of the selected configuration, it is evaluated on two additional datasets with different characteristics.

\begin{table*}[t]
    \caption{OCSVM Performance Across Different Datasets}
    \label{tab:dataset_optim_results}
    \centering
    \footnotesize
    \begin{tabular}{| l | c c c | c c c | c c c |}
        \hline
            \multirow{2}{*}{\textbf{Embedding}} &
            \multicolumn{3}{ c |}{\textbf{DRUPAL}} &
            \multicolumn{3}{ c |}{\textbf{SR-BH 2020}} &
            \multicolumn{3}{ c |}{\textbf{CSIC}} \\
            \cline{2-10}
             & 
            \textbf{FPR (\%) } & \textbf{TPR (\%)} & \textbf{AUC} &
            \textbf{FPR (\%)} & \textbf{TPR (\%)} & \textbf{AUC} &
            \textbf{FPR (\%)} & \textbf{TPR (\%)} & \textbf{AUC} \\
        \hline
              Word2Vec & 0.36 & 71.40 & 0.88 
               & 5.1 & 85.95 & 0.82 
               & 41.66 & 91.43 & 0.74 \\
        \hline
             FastText & \textbf{0.79} & \textbf{80.11} & 0.96
             & \textbf{3.4} & \textbf{90.90} & 0.92
             & \textbf{3.07} & \textbf{62.99} & 0.79  \\
         \hline
             Doc2Vec & 0.39 & 55.68 & 0.87 
             & 24.57 & 87.22 & 0.76
             & 16.70 & 72.39 & 0.82 \\  
           \hline    
    \end{tabular}
\end{table*}

The additional datasets considered are CSIC 2010 and SR-BH 2020. CSIC 2010 contains synthetically generated HTTP requests labeled as valid or malicious and has been widely used as a benchmark in previous studies \cite{CSIC2010Dataset}. In contrast, SR-BH 2020 comprises real-world traffic captured using honeypots and includes 12 distinct attack categories, enabling the model to be evaluated in a more realistic heterogeneous environment \cite{riera2022new}. Table~\ref{tab:dataset} summarizes the data distribution for both datasets.

Table~\ref{tab:dataset_optim_results} presents the performance of the OCSVM across the different datasets using the selected embedding configuration. The results in Table~\ref{tab:dataset_optim_results} reveal heterogeneous behavior of the OCSVM depending on the dataset. In Drupal, FastText achieves the best overall performance, combining a low FPR with the highest AUC and TPR, consistent with the findings from the embedding selection stage. In SR-BH 2020, although all representations achieve high TPR values, FastText maintains a more favorable balance between TPR and FPR, while Doc2Vec shows a substantial increase in FPR. Conversely, in CSIC 2010, a general loss of stability is observed, particularly for Word2Vec and Doc2Vec, which exhibit elevated FPR, suggesting a strong dependence on the dataset's synthetic nature. In summary, these results indicate that FastText provides more consistent performance across heterogeneous scenarios, while Word2Vec and Doc2Vec representations are more sensitive to changes in the underlying data distribution.

\subsection{Selection of One-Class Anomaly Detection Models}
After selecting the static embedding configuration that best represents the analyzed requests, a second experiment is conducted to evaluate different one-class anomaly detection models within the same embedding space. This experiment assesses the performance of multiple detection approaches and enables a fair comparison among detection algorithms under identical representational conditions.

Two additional one-class anomaly detection algorithms, namely Local Outlier Factor (LOF) and Isolation Forest (IF), are compared with the previously evaluated One-Class SVM (OCSVM). All models operate on fixed-length feature vectors derived from the selected static embedding representation. Each detection model is trained offline using only valid HTTP requests, following the same preprocessing pipeline and data partitioning strategy adopted in the previous experimental stage.

\begin{table*}[t]
    \caption{Performance Comparison of Embedding Representations Using Different One-Class Anomaly Detection Models}
    \label{tab:datasets-sidebyside}
    \centering
    \footnotesize
    \begin{tabular}{| l | l | c c c | c c c | c c c |}
        \hline
            \multirow{2}{*}{\textbf{Embedding}} &
            \multirow{2}{*}{\textbf{Algorithm}} &
            \multicolumn{3}{ c |}{\textbf{DRUPAL}} &
            \multicolumn{3}{ c |}{\textbf{SR-BH 2020}} &
            \multicolumn{3}{ c |}{\textbf{CSIC}} \\
            \cline{3-11}
             &  &
            \textbf{TPR (\%) } & \textbf{FPR (\%)} & \textbf{AUC} &
            \textbf{TPR (\%)} & \textbf{FPR (\%)} & \textbf{AUC} &
            \textbf{TPR (\%)} & \textbf{FPR (\%)} & \textbf{AUC} \\
        \hline
            \multirow{3}{*}{\textbf{FastText}}
            & OCSVM & \textbf{80.11} & \textbf{0.79} & 0.96 & \textbf{90.09} & \textbf{3.43} & 0.92 & 62.00 & 3.07 & 0.79 \\
            & LOF & 12.31 & 2.28 & 0.87 & 91.02 & 11.09 & 0.94 &\textbf{86.94} & \textbf{3.69} & 0.96 \\
            & IF  & 39.01 & 0.40 & 0.83 & 76.53 & 25.05 & 0.84 & 62.52 & 17.38 & 0.81 \\
        \hline
            \multirow{3}{*}{\textbf{Word2Vec}} 
            & OCSVM & 71.40 & 0.36 & 0.88 
            & 85.95 & 5.10 & 0.88
            & 91.43 & 41.66 & 0.74 \\

            & LOF & 7.38  & 2.97 & 0.86 
            & 91.97 & 11.53 & 0.93
            & 60.39 & 0.11 & 0.67 \\

            & IF  & 0.18  & 0.01 & 0.86 
            & 47.71 & 16.92 & 0.76 
            & 98.22 & 47.22 & 0.76 \\
          \hline

          \hline
            \multirow{3}{*}{\textbf{Doc2Vec}} 
             &  OCSVM & 54.73 & 0.39 & 0.87 
             & 87.22 & 24.57 & 0.76
             & 72.96 & 16.70 & 0.81
             \\

             & LOF & 44.69 & 0.23 & 0.89 
             & 56.57 & 6.23 & 0.81 
             & 79.94 & 18.93 & 0.87 \\
             
              & IF  & 54.92 & 0.50 & 0.90 
               & 98.27 & 29.83 & 0.91
               & 72.59 & 17.70 & 0.82 \\
           \hline
    \end{tabular}
\end{table*}

For each detector, a predefined set of hyperparameters is explored. In the case of LOF, the number of neighbors is varied with $k \in \{10, 20, 30, 50, 100, 150\}$. For Isolation Forest, the number of trees is selected from $n \in \{50, 100, 150, 200\}$. In both cases, the previously described operating point selection process is applied to calibrate the decision thresholds.

From the perspective of representation-based comparison, the results reported in Table~\ref{tab:datasets-sidebyside} reveal clear differences in the behavior of the evaluated models. FastText exhibits the most consistent performance across datasets, by achieving high AUC and TPR, particularly when combined with One-Class SVM, while maintaining relatively low FPR even in more complex scenarios such as SR-BH 2020. In contrast, Word2Vec exhibits more variable behavior, achieving satisfactory detection performance on some datasets but showing significant FPR degradation in scenarios such as CSIC 2010, suggesting greater sensitivity to changes in the data distribution. Doc2Vec, on the other hand, yields less stable results, with performance strongly dependent on the selected detection algorithm and a tendency to compromise the balance between detection capability and false-positive control. In summary, these findings indicates that the choice of the representation model has a considerable impact on the performance of one-class anomaly detection systems.

The results presented in Table~\ref{tab:baselines-results} further confirm that representation learning plays a critical role in anomaly detection performance across heterogeneous HTTP attack datasets. In the Drupal dataset, FastText-OCSVM achieves a balanced trade-off between detection capability and low FPR, significantly reducing the FPR compared to ModSecurity configurations and outperforming expert-designed baselines in terms of robustness. Although expert-selected features achieve slightly higher TPR, they incur higher FPR, which may limit their practical deployment in production environments.
\begin{table}[!t]
    \caption{Comparison of FastText-OCSVM Performance Against Baselines and Previous Work}
    \label{tab:baselines-results}
    \centering
    \footnotesize
    \begin{tabular}{| l | c | c |}
        \hline
            \multicolumn{3}{| c |}{\textbf{DRUPAL}} \\
        \hline
             \textbf{Approach} & \textbf{TPR (\%)} & \textbf{FPR (\%)} \\
        \hline        
         ModSecurity OWASP CRS v3 - PL1 & 29.55 & 15.57 \\
         ModSecurity OWASP CRS v3 - PL2 & 77.89 & 49.93 \\
         Expert-Selected features-OCSVM \cite{rmartinez-ladc-2018} & \textbf{94.40} & 6.00 \\
         Automatic-Selected features-OCSVM \cite{riverol2024capturing} & 91.76 & 2.29 \\
         FastText-OCSVM & 80.11 & \textbf{0.79} \\
        \hline
            \multicolumn{3}{|c|}{\textbf{SR-BH 2020}} \\
        \hline
             \textbf{Approach} & \textbf{TPR (\%)} & \textbf{FPR (\%)} \\
        \hline   
         ModSecurity OWASP CRS v3 - PL1 & 26.62 & 0.00 \\
         ModSecurity OWASP CRS v3 - PL2 & 28.48 & 0.00 \\
         Two-phase MultiOutput CatBoost \cite{riera2022new} & 88.82 & -- \\
         Automatic-Selected features-OCSVM \cite{riverol2024capturing} & 78.87 & \textbf{0.84} \\
         FastText-OCSVM & \textbf{90.09} & 3.43 \\
        \hline
            \multicolumn{3}{|c|}{\textbf{CSIC 2010}} \\
        \hline
             \textbf{Approach} & \textbf{TPR (\%)} & \textbf{FPR (\%)} \\
        \hline   
         ModSecurity OWASP CRS v3 - PL2 & 28.48 & 0.00 \\
         Expert-Selected features-SVM \cite{rmartinez-ladc-2018} & 39.63 & 5.37 \\
         Automatic-Selected features-SVM \cite{riverol2024capturing} & 35.74 & \textbf{0.49} \\
         FastText-CAE \cite{yamada2024evaluation} & 84.60 & 10.06 \\
         FastText-LOF & \textbf{86.94} & 3.69 \\
        \hline
    \end{tabular}
\end{table}
In the more challenging SR-BH 2020 dataset, FastText-OCSVM achieves the highest TPR (90.09\%) while maintaining a controlled FPR (3.43\%), demonstrating stronger generalization under evolving attack patterns. Notably, rule-based approaches such as ModSecurity exhibit extremely low FPR but at the cost of severely limited detection capability, highlighting the inherent trade-off between strict rule enforcement and adaptive learning-based detection.

For the CSIC 2010 dataset, the comparative behavior remains consistent: representation-based approaches improve detection rates relative to manually engineered features while maintaining acceptable false-positive rates. Although the absolute TPR is lower in this dataset, FastText-OCSVM still provides a more balanced performance profile than both rule-based and traditional feature-engineering approaches.

Overall, these findings reinforce the hypothesis that the choice of embedding representation significantly influences the stability and effectiveness of one-class anomaly detection systems. In particular, FastText demonstrates greater robustness across datasets with varying structural complexity, suggesting that subword-level modeling contributes to generalization in web traffic anomaly detection.

\section{Conclusion}
The impact of static embedding representations on HTTP request anomaly detection models was evaluated. The FastText embedding model achieved the best balance between TPR and FPR across all datasets. FastText's ability to model subwords mitigated the OOV word problem, allowing the construction of a canonical request representation base. This base allows the generation of representations for any input, achieving 100\% attack coverage compared to 6.6\% for other techniques. Among the algorithms evaluated, the OCSVM classifier showed the most stable behavior, balancing sensitivity and accuracy in all experiments. The combination of FastText embeddings with OCSVM was the most effective, reducing false positives without affecting detection and achieving the objectives. The HEDA implementation with FastText and OCSVM performs comparably to previous methods and surpasses the baseline, showing clear improvement over ModSecurity.

\section{Future work}
Future work will explore several directions to improve the representation of HTTP requests and the performance of the proposed detection framework.  First, different aggregation strategies for FastText embeddings will be evaluated. In particular, the impact of techniques such as max pooling and weighted averaging will be analyzed in order to determine whether they produce richer vector representations than the current averaging approach. Second, sliding-window aggregation mechanisms will be investigated as an alternative to global averaging across all token vectors. This approach aims to preserve sequential information and local contextual relationships within HTTP request strings. Finally, future research will explore integrating embedding-based representations with validation mechanisms applied to HTTP headers, combining semantic information from token embeddings with structural features extracted from request metadata within a unified detection framework.

\bibliographystyle{IEEEtran}
\bibliography{wafintl}

\vfill
\end{document}